\documentclass[journal=nalefd,manuscript=article]{achemso}

\usepackage[version=3]{mhchem} % Formula subscripts using \ce{}
\usepackage[T1]{fontenc}       % Use modern font encodings
\usepackage{textcomp}
\usepackage{amsmath}

\author{Josep Ingla-Ayn\'es}
\email{J.Ingla.Aynes@rug.nl}
\affiliation{Physics of Nanodevices, Faculty of Mathematics and Natural Sciences, Zernike Institute for Advanced Materials, University of Groningen, Groningen, The Netherlands}
\author{Rick J. Meijerink}
\affiliation{Physics of Nanodevices, Faculty of Mathematics and Natural Sciences, Zernike Institute for Advanced Materials, University of Groningen, Groningen, The Netherlands}
\author{Bart J. van Wees}
\affiliation{Physics of Nanodevices, Faculty of Mathematics and Natural Sciences, Zernike Institute for Advanced Materials, University of Groningen, Groningen, The Netherlands}

\title[An \textsf{achemso} demo]
  {98\% directional guiding of spin currents with 90 micrometer relaxation length in bilayer graphene using carrier drift}

\keywords{Graphene, boron nitride, spin transport, drift}

\begin{document}

%\begin{tocentry}

%Some journals require a graphical entry for the Table of Contents.
%This should be laid out ``print ready'' so that the sizing of the

%Inside the \texttt{tocentry} environment, the font used is Helvetica
%8\,pt, as required by \emph{Journal of the American Chemical

%The surrounding frame is 9\,cm by 3.5\,cm, which is the maximum
%permitted for  \emph{Journal of the American Chemical Society}
%graphical table of content entries. The box will not resize if the
%content is too big: instead it will overflow the edge of the box.

%This box and the associated title will always be printed on a
%separate page at the end of the document.

%\end{tocentry}

\begin{abstract}
Electrical control of spin signals and long distance spin transport are major requirements in the field of spin electronics. Here we report the efficient guiding of spin currents at room temperature in high mobility hexagonal boron nitride encapsulated bilayer graphene using carrier drift. Our experiments, together with modelling, show that the spin relaxation length can be tuned from 2 to 88~\textmu m when applying a DC current of $\mp$40~\textmu A respectively. Our model predicts that, extending the range up to $\mathrm{I_{dc}}=\mp$150~\textmu A, the spin relaxation length can be tuned from 0.6 to 320~\textmu m respectively, indicating that spin relaxation lengths in the millimeter range are within scope in near future with moderate current densities. 
Our results also show that we are able to direct spin currents on either side of a spin injection contact. 98\% of the injected spins flow to the left when $\mathrm{I_{dc}}$=~-40~\textmu A and 65\% flow to the right when the drift current is reversed. Our model shows that, for $\mathrm{I_{dc}}=\mp$150~\textmu A the numbers reach 99.8\% and 95\% respectively showing the potential of carrier drift for spin-based logic operations and devices.
\end{abstract}

%\section{Introduction}

Propagation of spins has been traditionally studied using spin diffusion, which is a slow, non directional process that limits the range over which spins can be transported without losing the spin polarization. In contrast, transport induced by carrier drift allows for fast and directional propagation of spins enabling long distance spin transport \cite{FabianRevModPhys}. This effect relies on the fact that a charge current is associated with an in-plane electric field $\mathrm{E}$, causing carriers to drift with a velocity $\mathrm{v_d=\mu E}$ which is proportional to the electronic mobility $\mathrm{\mu}$ of the channel. As a result, when a spin accumulation is present, the propagation of spins can be controlled with a drift field\cite{YuFlatte, IvanRapCom}.
Low temperature spin drift experiments performed in semiconductors such as silicon \cite{SpinTransportSiNature, SpinDriftDopedNtypeSi,RoomTemperatureHanleDriftSi} and gallium arsenide \cite{SpinDriftGaAs} showed a modulation of the Hanle spin precession with the applied bias. Room temperature modulation of the spin relaxation length between 0.85 and 4.53~\textmu m was obtained for Si \cite{RoomTemperatureHanleDriftSi}.

Graphene is a 2D material that presents outstanding electronic properties \cite{RevModPhysGraph, MobGeimPRL} and long spin relaxation times \cite{NikoSpinGraf, SarojCVD, 12nsAachen} that are ideal for spintronic applications \cite{RevRocheVal, ReviewFabian}.  Graphene's unprecedentedly high electronic mobilities $\mathrm{\mu}$ are an attractive incentive for spin drift measurements. Ref. \cite{DriftGraphenePRL} represents the proof of principle for this effect in graphene on SiO$_2$ at room temperature. However, the efficiency was limited by the low mobility and short spin relaxation time of the graphene samples on SiO$_2$. In the past years, several approaches have been used to enhance the electronic quality of graphene. In particular, the use of hexagonal boron nitride (hBN) as a substrate has lead to a great improvement of the graphene quality in terms of charge \cite{KimBN,1Dcontact} and spin transport \cite{PRLMarcos,AachenNanoletters,RapeCom,BLGBarbaros}. In this letter we show that the magnitude of the spin signal can be controlled efficiently by applying drift currents in high mobility hBN encapsulated bilayer graphene (BLG). Our results, together with a model that accounts for drift in our geometry, show that we have achieved a strong modulation of the spin relaxation length from 2 to 88~\textmu m when applying a moderate DC drift current($\mathrm{I_{dc}}$) of $\pm$40~\textmu A. Because of the agreement between the measured data and the model, we extend our analysis up to $\mathrm{I_{dc}}$=~$\pm$ 150~\textmu A. In this case, the spin relaxation length changes from 0.6 to 320~\textmu m, an almost 3 orders of magnitude modulation.

Also we demonstrate the efficiency of a drift field in directing the spin currents. Showing that we can steer the injected spin currents to the right and left sides of the injecting contact with efficiencies of 65\% and 98\% applying drift currents of $\pm$40~\textmu A respectively. 
%\section{Results and discussion}

When applying a drift field in the graphene channel, the spin accumulation follows the drift diffusion equation:
\begin{equation}
\mathrm{D_s} \frac {d^2n_s(x)}{dx^2}- v_d \frac {dn_s(x)}{dx}-\frac{n_s(x)}{\tau_s}=0 
\label{DriftDiff}
\end{equation}

Where $\mathrm{D_s}$ is the spin diffusion coefficient, $n_s$ the spin accumulation, $\tau_s$ the spin relaxation time and $v_d$=~$-(+)\mathrm{\mu E}$ when the carriers are electrons (holes).
As can be seen from Eq. 1, when an electric field is applied, the propagation of spin signals is no longer symmetric in the $\pm x$ direction.
This equation has solutions in the form of $n_s=\mathrm{A}\exp( x/\lambda_{+})+\mathrm{B}\exp(- x/\lambda_{-})$  where $\lambda_{+(-)}$ are the relaxation lengths for spins propagating towards the left (right) in our system (also called upstream (downstream) in the literature \cite{YuFlatte}).

\begin{equation}
\frac{1}{\lambda_{\pm}}=\pm\frac{v_d}{2D_s}+\sqrt{\left(\frac{1}{\sqrt{\tau_s D_s}}\right)^2+\left(\frac{v_d}{2D_s}\right)^2 }
\end{equation}
This asymmetry in the spin propagation allows us to direct the spin currents in a controlled way. Because $E=\mathrm{I_{dc}R_{sq}/W}$ where $\mathrm{R_{sq}}$ is the square resistance  and $\mathrm{W}$ the width of the channel, such control can be achieved using $\mathrm{I_{dc}}$ with an efficiency that is given by the applied electric field and the mobility of the device.\\

\begin{figure}
		\includegraphics[width=\textwidth]{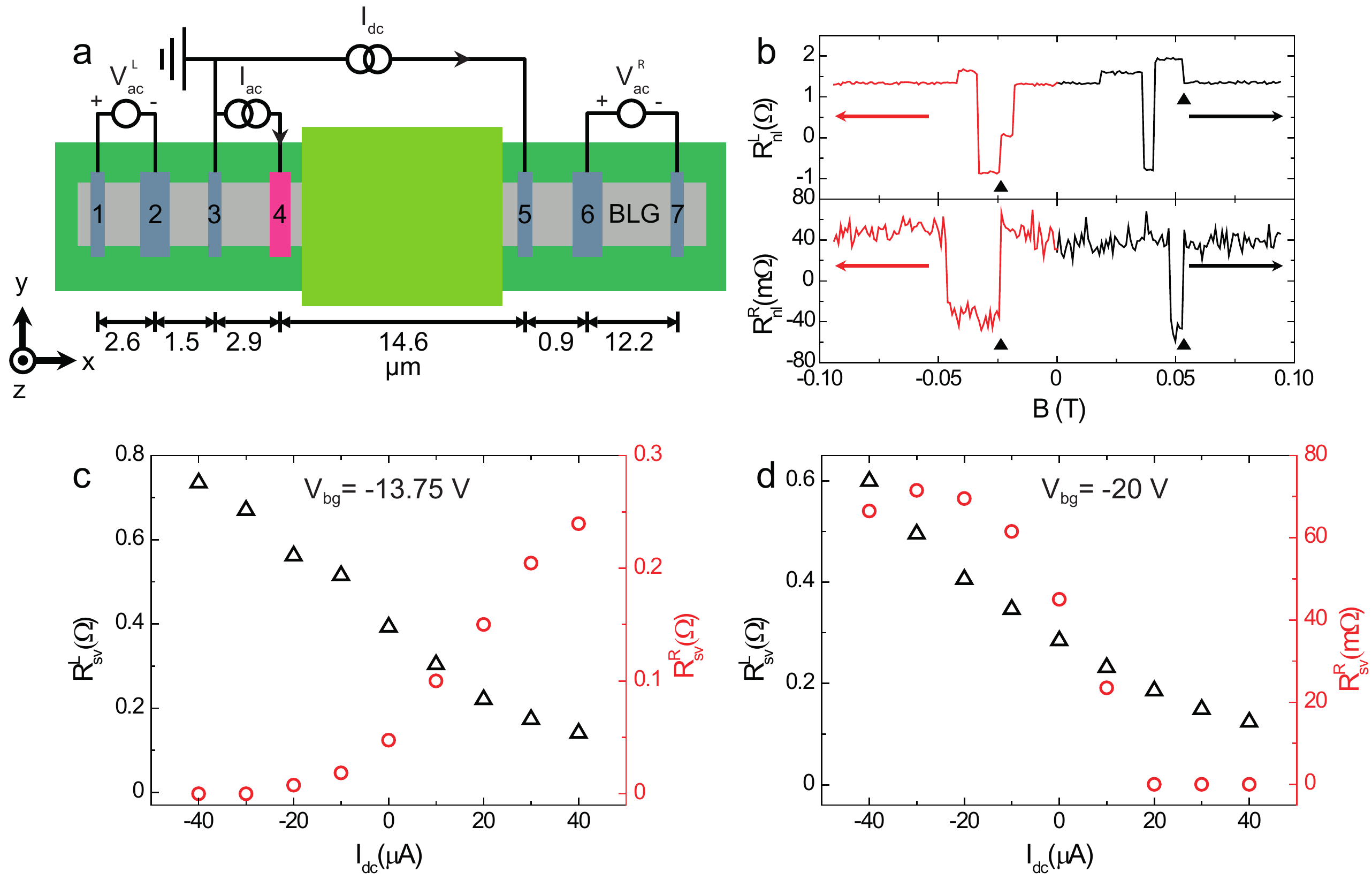}
	\caption{\textbf{a} Measurement geometry. The hBN/bilayer graphene/hBN stack is placed on a $n^{++}$ doped Si/SiO$_2$ substrate and the ferromagnetic contacts are made on the nonencapsulated regions. An AC current ($\mathrm{I_{ac}}$) is sent between contacts 4 and 3 to create a spin imbalance and a DC current ($\mathrm{I_{dc}}$) is sent between 5 and 3 to induce drift. The signal is detected simultaneously in left side of the injection circuit and across the encapsulated region ($\mathrm{V_{ac}^{L(R)}}$ respectively). The spacings between contacts are in \textmu m. \textbf{b} Nonlocal resistances in the left and right detection circuits (top and bottom panels respectively) as a function of an in-plane magnetic field at $\mathrm{I_{dc}}=$~0~\textmu A. Backgrounds of -5.5 and 4.85~$\Omega$ respectively have been substracted for clarity. The arrows indicate the magnetic field sweep direction and the triangles the switches caused by the magnetization reversal of contact 4. \textbf{c} and \textbf{d} Spin signal obtained in the left and right detectors (black triangles and red dots) as a function of $\mathrm{I_{dc}}$ at $\mathrm{V_{bg}}=$~-13.75~V where the carriers in the encapsulated region are electrons and at $\mathrm{V_{bg}}=$~-20~V where the carriers of the encapsulated region are holes respectively. In the outer regions the spins are carried by electrons in both cases.}
	\label{fig1}
\end{figure}

Our results are obtained using a bilayer graphene device that is partially encapsulated between two hBN flakes in the geometry shown in Fig. \ref{fig1}\textbf{a} and prepared using a dry transfer technique \cite{1Dcontact, FastPickUp}. The bilayer graphene obtained by exfoliation is supported by a bottom hBN flake (23~nm thick) and the ferromagnetic Co contacts (0.8~nm TiO$_x$/65~nm Co/5~nm Al) with widths ranging from 0.15 to 0.55~\textmu m are placed on the outer regions. The central region is encapsulated between both bottom and top hBN (21~nm thick) that is covered by a top gate (not shown for clarity) which we have set to zero voltage relative to contact 4. The spin and charge transport properties of this sample at room temperature and 4~K  can be found in \cite{RapeCom}.\\
We send an AC current ($\mathrm{I_{ac}}$) of 1 \textmu A between contacts 4 and 3 to inject spins. The in-plane electric field is applied by sending $\mathrm{I_{dc}}$ between contacts 5 and 3. We have used the standard low frequency lock-in technique to detect the AC spin signals (13~Hz) between contacts 2 and 1 and 6 and 7 simultaneously to study the effect of a drift current on the spin signal. When applying a magnetic field in the y direction, the contact magnetizations are controlled independently due to their different width that gives rise to different coercive fields. The results for the nonlocal spin signals are shown in  Fig. \ref{fig1}\textbf{b}. The nonlocal voltage is normalized by $\mathrm{I_{ac}}$ to obtain the nonlocal resistance in the left (right) side of the injector: $\mathrm{R_{nl}^{L(R)}=V_{ac}^{L(R)}/I_{ac}}$ where $\mathrm{V_{ac}^{L(R)}}$ is the voltage measured between contacts 2 and 1 (6 and 7).\\
In Fig. \ref{fig1}\textbf{b}, at $\mathrm{B}\approx$~-25~mT and $55$~mT we see simultaneous switching in $\mathrm{R_{nl}^R}$ and $\mathrm{R_{nl}^L}$ indicated with black triangles. Because no other switches occur simultaneously in both measurements, we attribute these switches to contact 4 that is our spin injector contact of interest \bibnote{Evolution of the switches with drift confirm that the switches are created by contact 4 and not by 3}.
 We define the spin signal: $\mathrm{R_{sv}^{L(R)}=\Delta R_{nl}/2}$ where $\mathrm{\Delta R_{nl}^{L(R)}}$ is the change in the nonlocal resistance in the left (right) detector caused by a switch of contact 4.\\
The carrier density of the BLG can be modified using the backgate \cite{GeimNovel}, formed in our case by the $n^{++}$ doped Si substrate and the 300~nm thick SiO$_2$ and 23~nm thick hBN gate insulators. 
In Fig \ref{fig1}\textbf{c} and \textbf{d} we show the spin signal dependence on the drift current at two different gate voltages (-13.75~V and -20~V respectively) corresponding to carrier densities of 3.3$\times10^{11}$~cm$^{-2}$ and -2.1$\times10^{11}$~cm$^{-2}$ in the encapsulated regions. These are chosen to obtain the largest drift velocity in the encapsulated region at a given $\mathrm{I_{dc}}$ (See supporting information (SI)). The outer regions are highly doped and the charge neutrality point is around -50~V, hence the carriers are electrons at both gate voltages.\\
When the encapsulated and nonencapsulated regions are both electron doped the spin signals measured at both detectors show an opposite trend with respect to $\mathrm{I_{dc}}$ (Fig. \ref{fig1}\textbf{c}). This can be understood taking into account that the detectors are at opposite sides of the injector contact and the carriers (electrons in both regions) are pushed towards the right (left) for positive (negative) drift velocities enhancing (reducing) the spin signal in the righ (left) detector. The control of the spin signal across the encapsulated region (right detector) is very efficient: At $\mathrm{I_{dc}}<$~-20~\textmu A the spin signal is supressed below the noise level (5~m$\Omega$) while at $\mathrm{I_{dc}= 40}$~\textmu A it is enhanced by 400\%. In the left detector, the modulation is dominated by the drift in the nonencapsulated region and we see that the spin signal is increased by 87\% for $\mathrm{I_{dc}}=$~-40~\textmu A  and it is reduced by 64\% when $\mathrm{I_{dc}= 40}$~\textmu A.\\
In Fig. \ref{fig1}\textbf{d} the carriers in the inner and outer regions have opposite polarity. In this case, the spin signals at both sides of the injector increase for negative $\mathrm{I_{dc}}$. This is because electrons and holes react in opposite ways when an electric field is created by $\mathrm{I_{dc}}$. In this case, the modulation of the spin signal across the encapsulated region is less efficient. It increases by 60\% for positive $\mathrm{I_{dc}}$ and it is suppressed below the noise level for $\mathrm{I_{dc}}<$~-20~\textmu A. We explain the smaller increase taking into account that, in this configuration and when applying a negative $\mathrm{I_{dc}}$, the electric field pulls the spins away from the injector (contact 4) in both directions and the spin accumulation below the injector decreases in a more pronounced way than at $\mathrm{V_{bg}}=$~-13.75~V. In Fig. \ref{fig1}\textbf{d}, we see that $\mathrm{R_{sv}^L}$ is slightly smaller than in Fig. \ref{fig1}\textbf{c}. However, in this case, the enhancement is more efficient than at $\mathrm{V_{bg}}=$~-13.75~V and the spin signal increases 100\% above its zero drift value. We attribute this to the higher resistance of the outer regions causing larger electric fields in the channel at the same $\mathrm{I_{dc}}$.

\begin{figure}
		\includegraphics[scale=0.5]{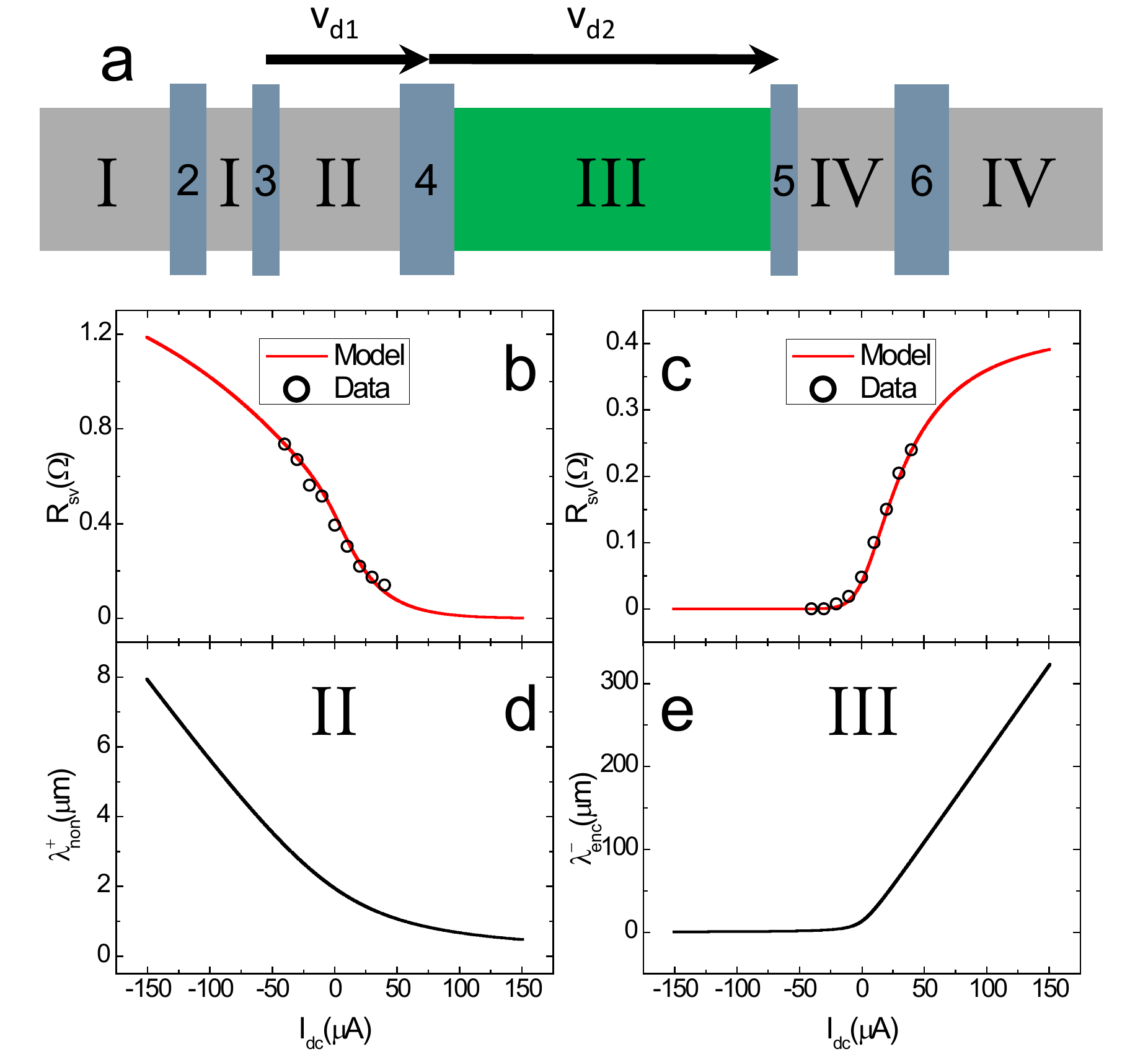}
	\caption{\textbf{a} Sketch of the simulated device with 4 different regions. I II and IV are the nonencapsulated regions and are assumed to have the same transport properties. III is the encapsulated region and drift is considered in regions II and III with drift velocities $\mathrm{v_{d1}}$ and $\mathrm{v_{d2}}$. The graphene is assumed to be infinite at both left and right sides. \textbf{b} and \textbf{c}: Amplitude of the spin signal generated by contact 4 in the detectors 2 and 6 respectively at $\mathrm{V_g}$=~-13.75~V. The red curves represent the nonlocal resistances obtained from the modelling. \textbf{d} Obtained relaxation length for spins propagating to the left in region II. \textbf{e} Obtained relaxation length for spins propagating to the right in region III (encapsulated region).}
	\label{fig2}
\end{figure}
To understand the the spin current distribution in the channel we have adapted the model developed in \cite{ElectronSpinPopinciuc} to the geometry shown in Fig. \ref{fig2}\textbf{a}. Region III (green) is encapsulated while the other ones are not. We account for the electric field applied in regions II and III using the drift diffusion equation (Eq. 1).\\

To extract the parameters needed for this model we have performed a similar analysis as in \cite{PRLMarcos, RapeCom}. The spin relaxation time in the nonencapsulated regions (I, II and IV in Fig. \ref{fig2}\textbf{a}) is extracted from Hanle precession measurements carried out in region I. The spin relaxation time in the encapsulated region is extracted using the 3 regions model derived in \cite{MarcosSuspended}. For this purpose, we have measured Hanle precession across the encapsulated region (III) and used the transport parameters of both encapsulated and nonencapsulated regions. The other parameters are extracted from the charge transport measurements (SI).\\ 

As shown in the SI, the agreement between the model and the experimental results, that is already good with the parameters mentioned above, can be improved by increasing the diffusion coefficient of the encapsulated region $\mathrm{D_{enc}}$ from 0.02 to 0.06~m$^2$/s and reducing the mobility $\mu_\mathrm{enc}$ from 2.8 to 2~m$^2$/(Vs). We notice that both changes reduce the effect of the drift and hence, our claims are not affected (SI).
We show the results taken at $\mathrm{V_g}=$~-13.75~V. In Fig. \ref{fig2}\textbf{b} and \textbf{c} we plot the experimental data shown in Fig \ref{fig1}\textbf{c} (dots) together with the values obtained from the modelling (line). There is a very good agreement between the model and the experimental values indicating the reliability of the model. Therefore we have extended the range of $\mathrm{I_{dc}}$ in the model to predict the effect of drift in a range that we did not explore experimentally to avoid breakdown of our ferromagnetic contacts. We see that the spin signal can be suppressed fully at both sides of the injection point, when applying a large enough current. Notice that the maximum value of $\pm$150~\textmu A is still one order of magnitude smaller than a typical breakdown current in bilayer graphene \cite{IbreakBLG} (Typically 1 mA for a 2~\textmu m wide device).\\
In Fig. \ref{fig2}\textbf{d} and \textbf{e} we show the spin relaxation lengths for spins diffusing towards the right (left) in the encapsulated (nonencapsulated) region ($\lambda_{non}^{+}$ and $\lambda_{enc}^{-}$ respectively). The difference observed between them is expected from the different transport properties of the encapsulated and nonencapsulated regions.
 
In the encapsulated region (Fig. \ref{fig2}\textbf{e}) the spin relaxation length increases up to 88 (320)~\textmu m for DC currents of 40 (150)~\textmu A when the spin relaxation length at zero DC current is 13~\textmu m. This observation shows the potential of drift currents to transport spins over long distances.\\

\begin{figure}
		\includegraphics[scale=0.5]{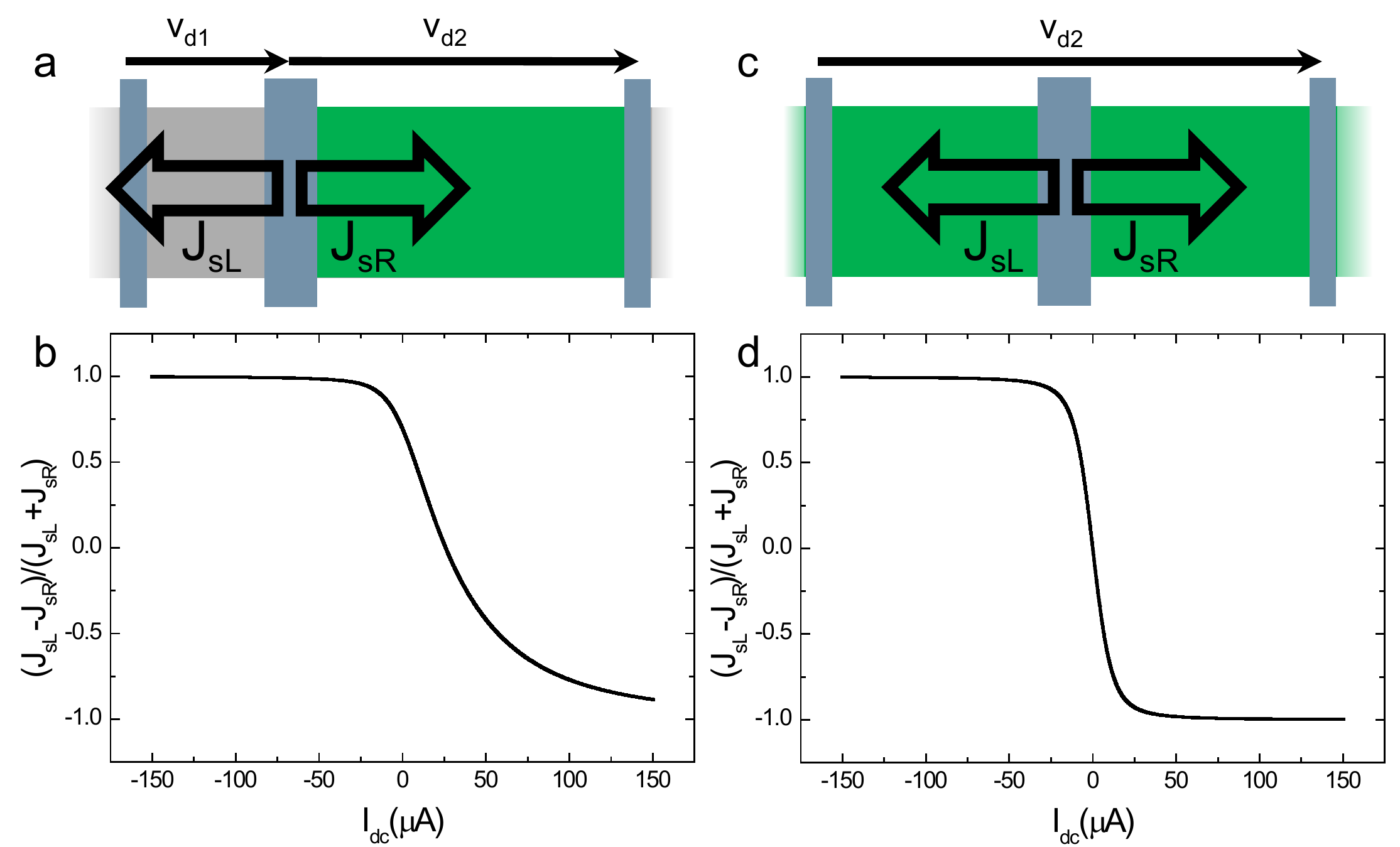}
	\caption{Directional control of spin currents. \textbf{a} Injected spin currents propagating towards the left (right) side of the injector ($\mathrm{J_{sL(R)}}$) in the modelled device geometry. \textbf{b} Directionality of the spin currents as a function of the DC current in our device extracted from the modelling discussed above. \textbf{c} Homogeneous geometry used to compare the results with a fully encapsulated sample where the spacing between the spin injector and the contacts where $\mathrm{I_{dc}}$  is applied is 14.6~\textmu m. \textbf{d} Calculated directionality of the spin currents as a function of $\mathrm{I_{dc}}$ for the geometry shown in \textbf{c}. The curve is symmetric in this case and higher efficiency in the modulation is achieved.} 
	\label{fig3}
\end{figure}

To account for future applications we have also studied if it is possible to direct spins in a specific direction. We define the directionality of the spin current as: $\mathrm{D=(J_{sL}-J_{sR})/(J_{sL}+J_{sR})}$ where $\mathrm{J_{sR(L)}}$ is the spin current towards the right (left) of the spin injector (contact 4). In Fig. \ref{fig3}\textbf{a} and \textbf{b} we show the $\mathrm{I_{dc}}$ dependence of $\mathrm{D}$ for our device geometry obtained from modelling. We see that, within our range, D is already controlled quite efficiently.  When $\mathrm{I_{dc}}$=~-40~\textmu A 98\% of the spins are directed to the left (D~=~0.97). When reversing the drift current, 65\% of the spins are guided towards the right (D~=~-0.3). The asymmetry with $\mathrm{I_{dc}}$ is caused by the different spin transport properties of both regions.
When extending the drift current range, the effect becomes very efficient. The model predicts that when applying a drift current of $\pm150$~\textmu A 95\% of the spins can be pushed towards the encapsulated region (D~=~0.9) while 99.9\% of them can be pushed towards the outer region (D~=~0.998).

To obtain the efficiency of this effect in a fully encapsulated device, we have calculated the $\mathrm{D}$ parameter in the homogeneous geometry shown in Fig. \ref{fig3}\textbf{c}. The results are shown in \textbf{d} and are symmetric with respect to $\mathrm{I_{dc}}$ as expected. From this calculation we can see that 99\% of the injected spins can be sent to either direction applying $\mathrm{I_{dc}}$~=~$\pm$40~\textmu A and, when $\mathrm{I_{dc}}$~=~$\pm$150~\textmu A, the effect is increased up to 99.9\%. Such efficiency is caused by the large drift velocities of 1.2$\times$10$^5$~m/s induced in the encapsulated region of our sample at $\mathrm{I_{dc}}$~=~150~\textmu A.

%\section{Conclusions}
In conclusion, we have shown that the spin transport can be controlled efficiently by applying drift currents in high mobility hBN encapsulated bilayer graphene. Our results, together with a model, show that we have achieved a strong modulation of the spin relaxation length from 2 to 88~\textmu m when applying a drift current of $\pm$40~\textmu A. Extending our analysis up to $\mathrm{I_{dc}}$=~$\pm$ 150~\textmu A we see that the spin relaxation length changes from 0.6 to 320~\textmu m, an almost 3 orders of magnitude modulation suggesting that 2 millimeter spin relaxation lengths should be achievable for $I_{dc}=$~1~mA. 

We notice that we cannot explore the full potential of the spin drift because the length of the graphene channel in exfoliated devices is constrained by the size of the flakes which can be obtained. Recent advances obtaining ultrahigh quality CVD graphene \cite{AachenCVDPickUp, AachenCVD28umMFP} make it possible to obtain high quality large devices showing spin transport over unprecedentedly long distances.

Using our model we also extract the directionality of the spin currents. We find that, when a drift current of -40~\textmu A is applied, 98\% of the spins are directed towards the left. When applying a DC current of 40~\textmu A 65\% of the spins are directed to the right. These results show that we have achieved efficient directional control of the spin currents at room temperature. 
Extending our range to $\mp$150~\textmu A these numbers rise up to 99.8\% and 95\% respectively showing that the control we achieved of the directionality of the spin propagation can enable new types of spin-based logic operations.

The directional control of spin currents achieved in our experiment shows that it is possible to realize logic operations using spin currents in a material with low spin orbit coupling such as graphene, opening the way to new device geometries and functionalities.

%\section{Experimental}

%The usual experimental details should appear here.  This could
%include a table, which can be referenced as Table~\ref{tbl:example}.
%Notice that the caption is positioned at the top of the table.

\begin{acknowledgement}
The authors thank J. C. Leutenantsmeyer, I. J. Vera-Marun and M. H. D. Guimar\~es for insightful discussions and H. Adema, J. G. Holstein, H. M. de Roosz and T. Schouten for technical assistance.
The research leading to these results has received funding from the People Programme (Marie Curie Actions) of the European Union's Seventh Framework Programme FP7/2007-2013/ under REA grant agreement n$^{\circ}$607904-13 Spinograph and the European Union Seventh Framework Programme under grant agreement n$^{\circ}$604391 Graphene Flagship.

\end{acknowledgement}

\begin{suppinfo}
\subsection{Modelling parameters}

In this section we discuss the determination of the spin and charge transport parameters for the modelling shown in the manuscript and the results obtained from the model using the parameters extracted from Hanle precession and charge transport measurements.\\

To determine the square resistance and mobility of the bilayer graphene in the encapsulated and nonencapsulated regions we carried out standard 4 probe measurements of the channel resistance  varying the backgate voltage at zero topgate voltage (Fig. \ref{SHanle}\textbf{a} and \textbf{b}).	The mobilities ($\mu$) extracted from  this curve are 2.8~m$^2$/(Vs) for the encapsulated region and 0.56~m$^2$/(Vs) for the nonencapsulated one. These values were extracted by fitting the square resistance versus the carrier density in the channel $\mathrm{n}$ using the formula: $\mathrm{R_{sq}=1/(ne\mu+\sigma_0)+\rho_s} $ where $\mathrm{e}$ is the electron charge, $\mathrm{\sigma_0}$ accounts for the finite resistance at the charge neutrality point. $\rho_s$ is an offset resistance attributed to short range scattering \cite{MobGeimPRL}. In the encapsulated region the fitting was done in the range -6.2<$V_{bg}$<35~V to avoid the underestimation of the carrier density produced by electron-hole puddles close to the neutrality point and the obtained values are $\mathrm{\sigma_0}=$~-1.7$\times10^{-3}$~$\Omega^{-1}$ and $\rho_s=$~4.3~$\Omega$.\\

\begin{figure}
		\includegraphics[width=\textwidth]{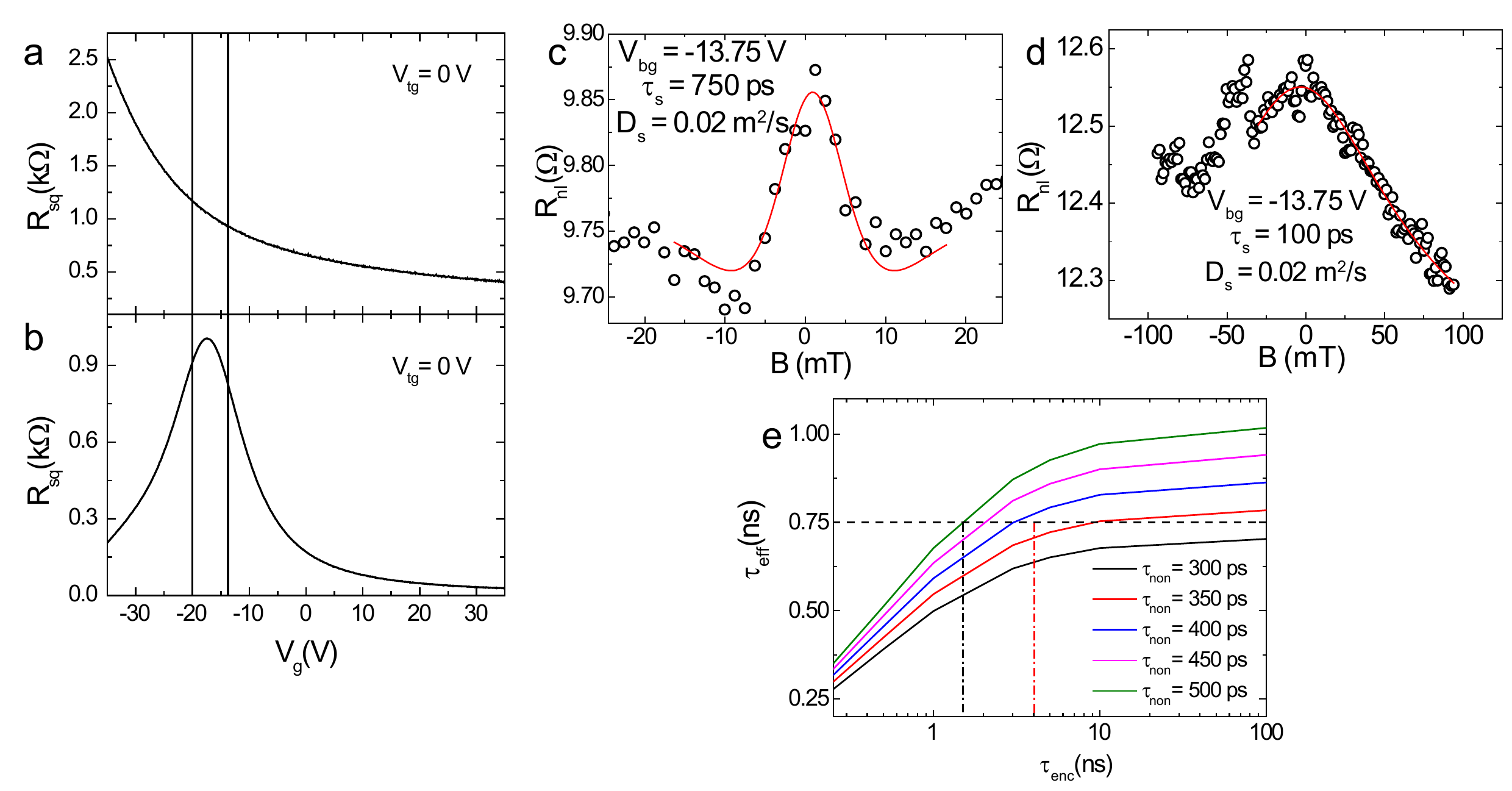}
	\caption{Charge and spin transport measurements carried out to characterize the encapsulated and nonencapsulated regions.\textbf{a} $\mathrm{R_{sq}}$ of the nonencapsulated and \textbf{b} encapsulated regions as a function of $\mathrm{V_{bg}}$. \textbf{c} and \textbf{d} Hanle precession curves obtained across the encapsulated and nonencapsulated regions respectively at $\mathrm{V_{bg}}=$~-13.75~V together with the fitting to the Bloch equations (red lines) and the corresponding parameters. \textbf{e} Effective spin relaxation times in the system $\mathrm{\tau_{eff}}$ as a function of the spin relaxation time in the encapsulated region $\mathrm{\tau_{enc}}$ for different spin relaxation times at the outer regions $\mathrm{\tau_{non}}$.}
	\label{SHanle}
\end{figure}
	
\begin{table}
 \caption{Spin and charge transport parameters at $V_{bg}=~-13.75$~V}
  \label{tbl:SpinParameters}
  \begin{tabular}{|c|c|c|c|c|c|c|c|}
   \hline
  $\mathrm{R_{enc}}$  &  $\mathrm{R_{non}}$   & $\mathrm{\mu_{enc}}$  & $\mathrm{\mu_{non}}$   & $\mathrm{D_{enc}}$ & $\mathrm{D_{non}}$  & $\mathrm{\tau_{enc}}$ & $\mathrm{\tau_{non}}$ \\
   \hline
    830~$\Omega$&  930~$\Omega$ &2.8~m$^2$/(Vs)&0.56~m$^2$/(Vs)   &0.02~m$^2$/s   &0.02~m$^2$/s  &3~ns  & 100~ps  \\
    
      \hline
    
  \end{tabular}
  \end{table}
\begin{table}
 \caption{Resistances of the contacts defined as in Fig. \ref{S1}. }
  \label{tbl:Rc}
  \begin{tabular}{|c|c|c|c|c|c|}
   \hline
  $\mathrm{R_{C2}}$  &  $\mathrm{R_{C3}}$ & $\mathrm{R_{C4}}$ & $\mathrm{R_{C5}}$ & $\mathrm{R_{C6}}$  \\
   \hline
820~$\Omega$ & 870~$\Omega$& 850~$\Omega$ & 1.48~k$\Omega$&1.56~k$\Omega$\\
      \hline
    
  \end{tabular}
\end{table}	
In Fig \ref{SHanle}\textbf{c} we show the Hanle precession curve measured across the encapsulated region. The amplitude of this signal is small due to the fact that in this regime the resistance of encapsulated and outer regions are comparable and the spin relaxation length of the encapsulated region is much longer than the one of the outer regions. This makes the spins diffuse and relax in the outer regions instead of crossing the encapsulated region, reducing the amplitude of the spin signal considerably. The asymmetry of the Hanle precession data with respect to zero magnetic field has been observed before \cite{AachenNanoletters, PRLMarcos, RapeCom} and we attribute it to a small misalignment of the contacts with respect to each other due to a poor adhesion with the bottom hBN. The Hanle precession gives us information about the spin relaxation time in the system but, in order to extract the spin transport properties of the encapsulated region, we have used a 3 regions model as in \cite{MarcosSuspended, PRLMarcos, RapeCom}. This model requires to determine the spin relaxation time of the outer regions. This is done by fitting the Hanle curve obtained in the nonencapsulated region and shown in Fig. \ref{SHanle}\textbf{d}. The shape change at negative magnetic fields is attributed to a switch in one of the contacts and this part is not included in the fit. Because of the short spacing between the contacts the shoulders characteristic of the Hanle curves are not present and we could not extract $\mathrm{D_s}$. As a consequence, we used the charge diffusion coefficient ($\mathrm{D_c}$). This is justified since in our devices there is an agreement between $\mathrm{D_c}$ and $\mathrm{D_s}$ \cite{RapeCom}. Because the contact resistances in our device are between 800~$\Omega$ and 1.5~k$\Omega$ (Table \ref{tbl:Rc}) the extracted spin relaxation time is reduced by the contacts and the spin relaxation time is a lower bound for the properties of the outer region\cite{ContactsHanle}.\\
In Fig. \ref{SHanle}\textbf{e} we show the effective spin relaxation time of our system as a function of the spin relaxation time in the encapsulated region for different values of the spin relaxation time in the nonencapsulated regions $ \mathrm{\tau_{out}}$. We notice that for $\mathrm{\tau_{out}}$ shorter than 350~ps $\mathrm{\tau_{eff}}$ is smaller than the measured value (dash line) even for $\mathrm{\tau_{enc}}$~=~100~ns. In order to get a reasonable estimate for the spin relaxation time in the encapsulated region we vary the spin relaxation time in the outer regions up to 500~ps. From these results we see that the spin relaxation time of the encapsulated region has to be longer than 1~ns and, to take a reasonable value not longer than the maximum spin relaxation time measured in this sample at 4~K \cite{RapeCom}, we take it 3~ns. 

The polarizations of the contacts used for the modelling were taken to assure good agreement between the model and the experimental data at zero drift current. The spin injection efficiency of the injector was set to 5\% while the obtained efficiency of the left and right detectors is $P_{d1}=$1.6\% and $P_{d2}=$7.5\% respectively to accomplish the abovementioned condition.

\subsection{Derivation of the Model}

In order to obtain the amplitude of the spin signal for different drift currents we have developed a model that uses the drift diffusion equations derived in \cite{YuFlatte}:
$$
D_s \nabla^2 n_s- v_d \nabla n_s-\frac{n_s}{\tau_s}=0 
$$
Here $D_s$ is the spin diffusion coefficient, $n_s$ the spin accumulation, $\mu$ the electronic mobility, $E$ the electric field and $\tau_s$ the spin relaxation time.
This equation has solutions in the form of $n_s=A\exp( x/\lambda_{+})+B\exp(- x/\lambda_{-})$ where $\lambda_{+(-)}$ correspond to the so called `upstream' and `downstream' spin relaxation lengths,
$$
\frac{1}{\lambda_{\pm}}=\pm\frac{v_d}{2D_s}+\sqrt{\left(\frac{1}{\sqrt{\tau_s D_s}}\right)^2+\left(\frac{v_d}{2D_s}\right)^2 }
$$
 where $v_d=\pm\mu E$ is the drift velocity and is negative for electrons and positive for holes, $\lambda_s=\sqrt{\tau_s D_s}$ is the spin relaxation length at zero drift.\\
 The spin current density $j_s$ is defined:
 $$
j_s(x) = -D_s \frac{d n_s (x)}{dx}+v_d n_s (x) 
 $$

\begin{figure}
		\includegraphics[width=\textwidth]{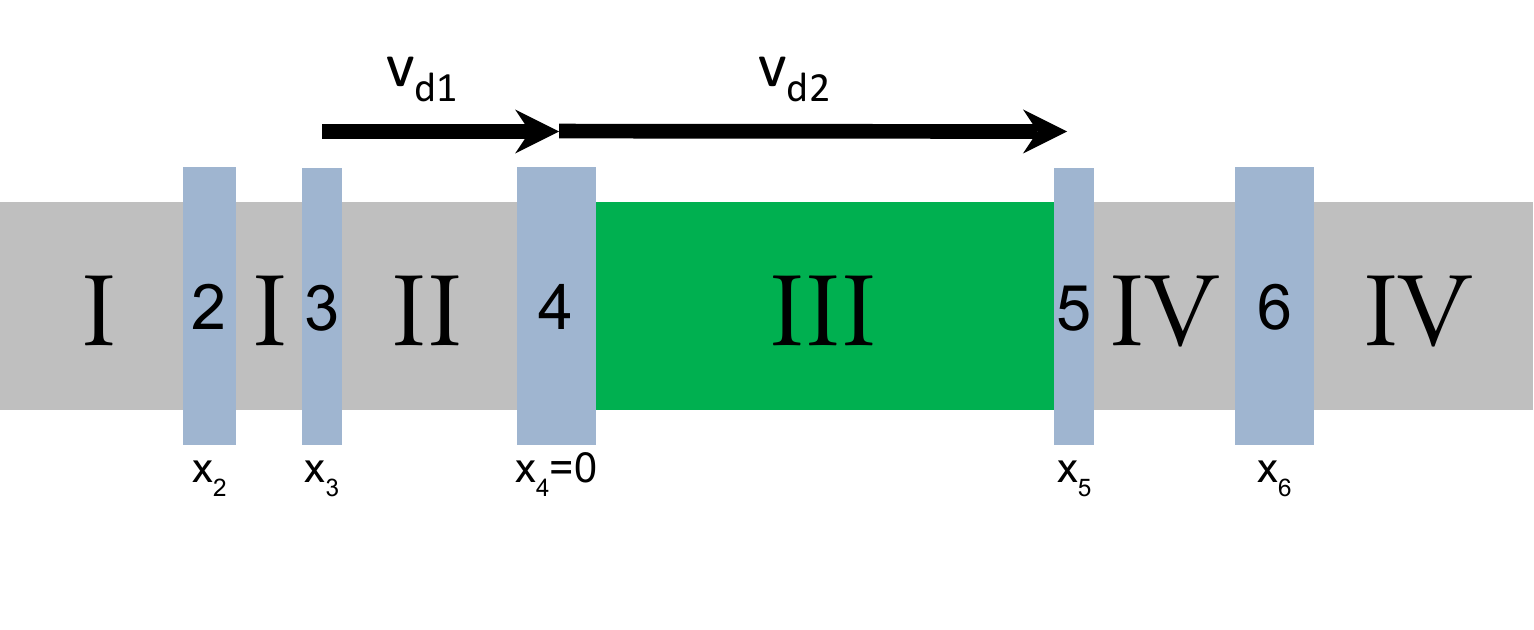}
	\caption{Sketch of the device geometry studied. Regions I, II and IV are non encapsulated while region III is encapsulated and hence has different properties. The electric field is only present in regions II and III. The DC current used to induce drift is sent between contacts 5 and 3 and the AC current used to inject spins is sent between contacts 4 and 3. Because of our analysis we do not consider any spin injection from contact 3. 2 and 6 are the detectors.}
	\label{S1}
\end{figure}

We write down the solution of the drift diffusion equations for the 4 different regions:
\begin{equation}
\begin{split}
\mathrm{I:}~~ n_s (x) &= A \exp(x/ \lambda _{non}) \\
\mathrm{II:}~~ n_s (x) &= B \exp(x/ \lambda _{non}^+)+C \exp(-x/\lambda _{non}^-)\\
\mathrm{III:}~~ n_s (x) &= D \exp(x/ \lambda _{enc}^+)+E \exp(-x/\lambda _{enc}^-)\\
\mathrm{IV:} ~~ n_s (x) &= F \exp(-x/ \lambda _{non})
\end{split}
\end{equation}
Here $\lambda _{(non)enc}$ are the spin relaxation lengths in the (non)encapsulated regions and $A-F$ are constants to be determined. 
We have used the boundary conditions $\mu_s (x\rightarrow\pm \infty)\rightarrow 0$. 

In order to obtain the spin signal in the geometry shown in \ref{S1} we apply the boundary conditions introduced in \cite{ElectronSpinPopinciuc} for the spin accumulation and spin currents at $\mathrm{x = x_3}$, $\mathrm{x_4=0}$ and $\mathrm{x_5}$.

The continuity of the spin accumulation reads:
\begin{equation}
\begin{split}
\mathrm{x= x_3:}&~~A\exp(x_3/ \lambda _{non})= B \exp(x_3/ \lambda _{non}^+)+C \exp(-x_3/\lambda _{non}^-)\\
\mathrm{x= 0:}&~~B+C = D+E\\
\mathrm{x= x_5:}&~~D \exp(x_5/ \lambda _{enc}^+)+E \exp(-x_5/\lambda _{enc}^-) = F \exp(-x_5/ \lambda _{non})
\end{split}
\label{Continuityns}
\end{equation}

And the continuity of the spin currents:

\begin{equation}
\begin{split}
\mathrm{x= x_3:}~~&A \frac{D_{non}}{\lambda_{non}}\exp(x_3/\lambda_{non})-B\left( \frac{D_{non}}{\lambda_{non}^+}-v_d^{non}\right)\exp(x_3/\lambda_{non}^+)\\
+&C \left( \frac{D_{non}}{\lambda_{non}^-}+v_d^{non}\right)\exp(-x_3/\lambda_{non}^-) = 0\\
\mathrm{x= 0:}~~&B\left( \frac{D_{non}}{\lambda_{non}^+}-v_d^{non}\right)-C \left( \frac{D_{non}}{\lambda_{non}^-}+v_d^{non}\right)\\
-&D\left( \frac{D_{enc}}{\lambda_{enc}^+}-v_d^{enc}\right)+E\left( \frac{D_{enc}}{\lambda_{enc}^-}+v_d^{enc}\right)=\frac{P_i I_{ac}}{W} \\
\mathrm{x= x_5:}~~&D\left( \frac{D_{enc}}{\lambda_{enc}^+}-v_d^{enc}\right)\exp(x_5/\lambda_{enc}^+)-E\left( \frac{D_{enc}}{\lambda_{enc}^-}+v_d^{enc}\right)\exp(-x_5/\lambda_{enc}^-)\\
+&F \frac{D_{non}}{\lambda_{non}}\exp(-x_5/\lambda_{non})=0
\end{split}
\label{SpinCurrentsns}
\end{equation}
$D_{(non)enc}$, $R_{(non)enc}$ and $v_d^{(non)enc}$ are the spin diffusion coefficient, square resistance and drift velocity of the (non)encapsulated regions and $P_i$ is the spin polarization of contact 4.
From equations 4 and 5 we have 6 equations that are used to solve for the 6 unknown parameters $A-F$. To obtain the nonlocal resistances from $n_s(x)$ we need to derive the spin electrochemical potential $\mu_s(x)$. 
We use the Einstein relation to write  $\mu_s(x)=en_s(x)R_{sq}/D_c$,
where $D_c$ is the charge diffusion coefficient. The nonlocal resistance at the detectors can be now obtained:
\begin{equation}
R_{nl}=P_d\mu_s(L)/(eI_{ac})=P_d R_{sq}n_s(L)/(D_cI_{ac})
\end{equation}
We obtain the nonlocal resistance in contacts 2 and 6.
\begin{equation}
\begin{split}
2:~~ &R_{nl}= P_{d1} R_{non} A \exp(x_2/\lambda_{non})/(D_{non}I_{ac})\\
6:~~&R_{nl}= P_{d2} R_{non} F \exp(-x_6/\lambda_{non})/(D_{non}I_{ac})
\end{split}
\end{equation}
Where $P_{d1}$ and $P_{d2}$ are the polarization of the detectors in regions I and IV respectively, $R_{non}$ and $D_{non}$ are the sheet resistivity and the diffusion coefficient in the nonencapsulated region and $x_{2(6)}$ are the positions of the detectors as defined in \ref{S1}.

\subsection{Results from the model and discussion of the parameters}
\begin{figure}
		\includegraphics[width=\textwidth]{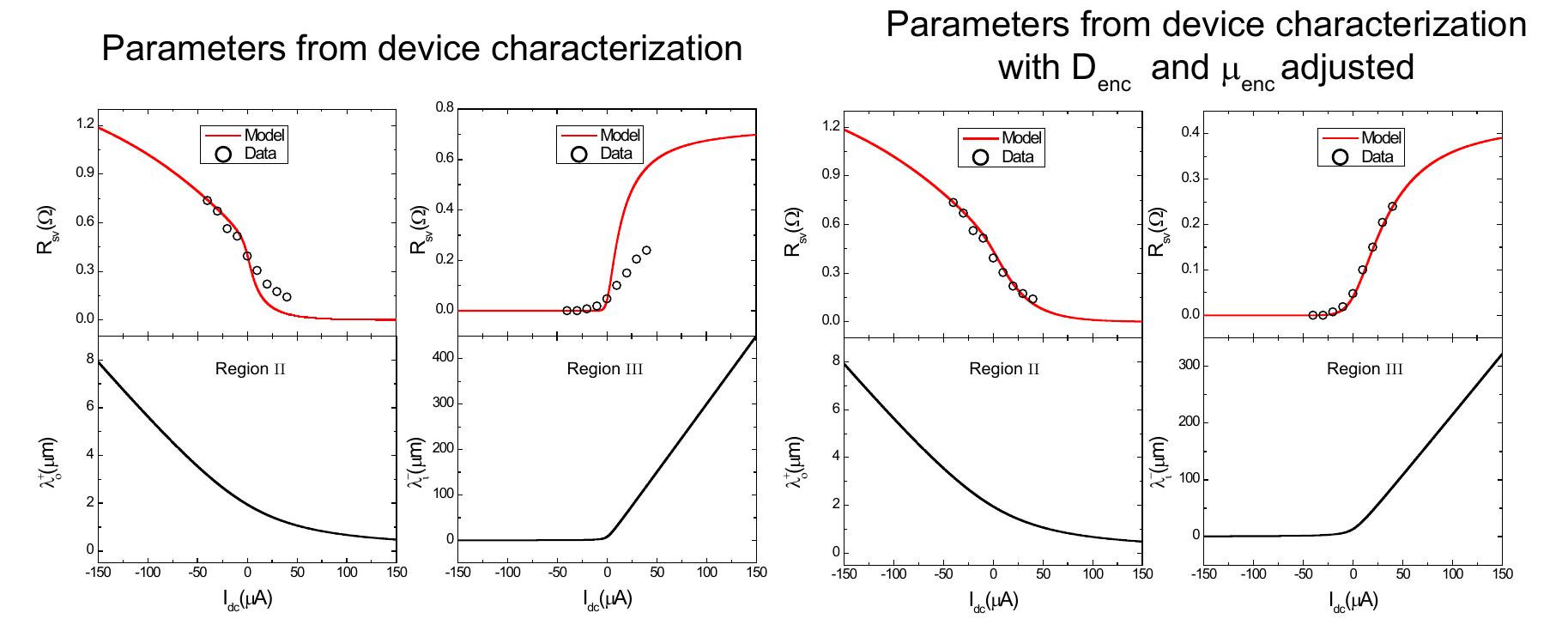}
	\caption{Comparison between the results of our modelling using the parameters extracted from device characterization (\textbf{a} and \textbf{b}) and the modelling obtained adapting $\mathrm{D_{enc}}$ and $\mathrm{\mu_{enc}}$ to improve the agreement (\textbf{c} and \textbf{d}). }
	\label{SParam}
\end{figure}

Using the parameters mentioned above, we used our drift model and compared the results with the experimental data as shown in Fig.\ref{SParam}\textbf{a} and \textbf{b}. As we can see from there, there is a reasonable agreement for both cases, which can be considered quite good taking into account the complex device geometry and the experimental uncertainties. We observe that the predicted effect (red lines) is stronger than the one measured experimentally and, in order to improve the agreement and extract more reliable conclusions, we reduced the mobility of the encapsulated region from 2.8 to 2~m$^2$/(Vs) and increased the diffusion coefficient of the encapsulated region from 0.2 to 0.6~m$^2$/s. This improves the agreement as shown in Fig.\ref{SParam}\textbf{c} and \textbf{d}. We notice that these changes do not affect our claims. Since the mobility is reduced  $\lambda_{enc}^-$ becomes lower at high drift values and, our claims are therefore a lower bound for the spin relaxation length in the encapsulated region for positive $\mathrm{I_{dc}}$.
Because by adapting the parameters we increase the diffusion coefficient of the encapsulated region, this also increases the asymmetry in the directional control of the spin current reducing the effect and, hence, our claims with respect to the directionality of spin currents in our device are also a lower bound of the real effect.
\end{suppinfo}
\bibliography{bibliography}

\begin{mcitethebibliography}{31}
\bibitem[Not()]{Note-1}
Evolution of the switches with drift confirm that the switches are created by
  contact 4 and not by 3\relax
\mciteBstWouldAddEndPuncttrue
\mciteSetBstMidEndSepPunct{\mcitedefaultmidpunct}
{\mcitedefaultendpunct}{\mcitedefaultseppunct}\relax
\EndOfBibitem
\end{mcitethebibliography}

\end{document}